\RequirePackage[2020/08/10]{latexrelease}
\documentclass[aps,pra,reprint,superscriptaddress,longbibliography]{revtex4-1}

\usepackage{physics}
\usepackage{amsmath}
\usepackage{amssymb}
\usepackage{graphicx}
\usepackage{float}
\usepackage[margin=0.7in]{geometry}
\graphicspath{{PNG/}}
\usepackage{xcolor}

\begin{document}

\title{Optical focusing of Bose-Einstein condensates}

\author{A.~M.~Kordbacheh}
\author{S.~S.~Szigeti}%
\affiliation{ 
	Department of Quantum Science, Research School of Physics, The Australian National University, Canberra, ACT 2601, Australia
}
\author{A.~M.~Martin}%
\affiliation{ 
	School of Physics, University of Melbourne, Melbourne, 3010, Australia
}%

\date{\today}

\begin{abstract}
We theoretically investigate the optical focusing of a rubidium Bose-Einstein condensate onto a planar surface. Our analysis uses a Gaussian variational method that includes the effects of two-body atom-atom interactions and three-body recombination losses. The essential factors such as the width, peak density and atom loss rate of the focused BEC profile on the surface are investigated and compared to Gross-Pitaevskii numerical simulations. We find a reasonable agreement in the results between our analytical approach and the numerical simulations. Our analysis predicts that condensates of $10^5$ atoms could be focused down to $\sim 10$nm widths, potentially allowing nanometer-scale atomic deposition with peak densities greater than $10^5$ atoms/$\mu$m$^2$.

\end{abstract}

\maketitle

\section{Introduction}

Atom lithography aims to deposit nanostructures onto a surface via the direct manipulation of cold-atom beams with optical fields \cite{1,2}. It potentially offers a controlled and flexible deposition procedure at the atomic scale, which could increase the density of transistors in computer chips \cite{3, 54}. Early theoretical works that detailed how optical beams could be used to focus atomic beams to nanoscale spot sizes  \cite{44, 45, 46} were soon followed by experimental demonstrations of direct depositions with sodium \cite{1} and chromium \cite{2} atoms. Direct atomic deposition has also been demonstrated with ytterbium \cite{50} and iron \cite{51,52}, and is capable of the precise generation of 2D and 3D nanostructures \cite{49}.

Almost all the experiments accomplished so far have used an oven source of atoms in which the beam is collimated with an aperture followed by a transverse laser cooling process \cite{2} before traveling through a focusing potential. However, there are advantages to using a Bose-Einstein condensate (BEC) of neutral atoms for atom deposition. Since the de Broglie wavelength of an atomic gas is on the order of the mean distance between particles, a BEC source would bring atoms to wavelengths between 1 nm and 1 pm for nano-Kelvin temperatures \cite{4}, resulting in an excellent collimation of the beam of atoms as well as a high flux density \cite{5}. Using a BEC source can also reduce effects such as chromatic aberration and angular divergence \cite{20}, with the longitudinal and transverse velocity distributions typically being much narrower when incident on the surface compared to those resulting from oven or thermal sources. Providing simple, quantitative estimates of the deposited focused widths and peak densities of a BEC source is the subject of this paper.

For thermal atomic sources, the atomic density is sufficiently dilute that interatomic interactions are negligible. Consequently, a classical approach based on single-atom trajectories is suitable in situations where the wave-like properties of the atoms are not too important, in analogy to ray optics \cite{n1}, with some wave-like effects such as diffraction and chromatic aberration accounted for in an \textit{ad hoc} manner \cite{n2, n3, n4}. However, the effect of interactions must be accounted for in BEC sources. Interatomic interactions in a BEC are dominated by \textit{s}-wave scattering \cite{10,11,12,13}, and in certain atomic species can be tuned from strongly repulsive to strong attractive via a Feshbach resonance \cite{29}. The matterwave focusing dynamics of trapped BECs in both repulsively interacting and non-interacting regimes have been investigated previously \cite{53}. The significance of our work is to take the atomic interaction into account when focusing a freely propagating BEC and to scale its effect on the broadening of the nano-focal spot sizes and peak densities achievable in realistic nano-lithography experiments with cold atoms. The other substantial point considered in this work is the effect of three-body recombination losses \cite{15,16,17} on the focusing scheme. This effect becomes especially striking in the high flux regimes of focus. The influence of three-body losses can be controlled via the sign and strength of the \textit{s}-wave interactions.

In this paper, we use an analytical variational approach based on a Gaussian ansatz \cite{14} to model the focusing dynamics of a freely falling BEC. Our model assumes a harmonic optical focusing potential, with optimal focal parameters determined via a classical particle trajectories approach \cite{20} and includes two-body atom-atom interactions and three-body recombination losses. We obtain estimates of the deposited focused BEC widths and peak densities. We also present corresponding Gross-Pitaevskii equation (GPE) numerical results to test the validity of the Gaussian ansatz approach.

\section{\label{sec:level0}Problem Description}

\begin{figure}[tb]
\centering
\includegraphics[width=9cm, height=8cm,angle=0]{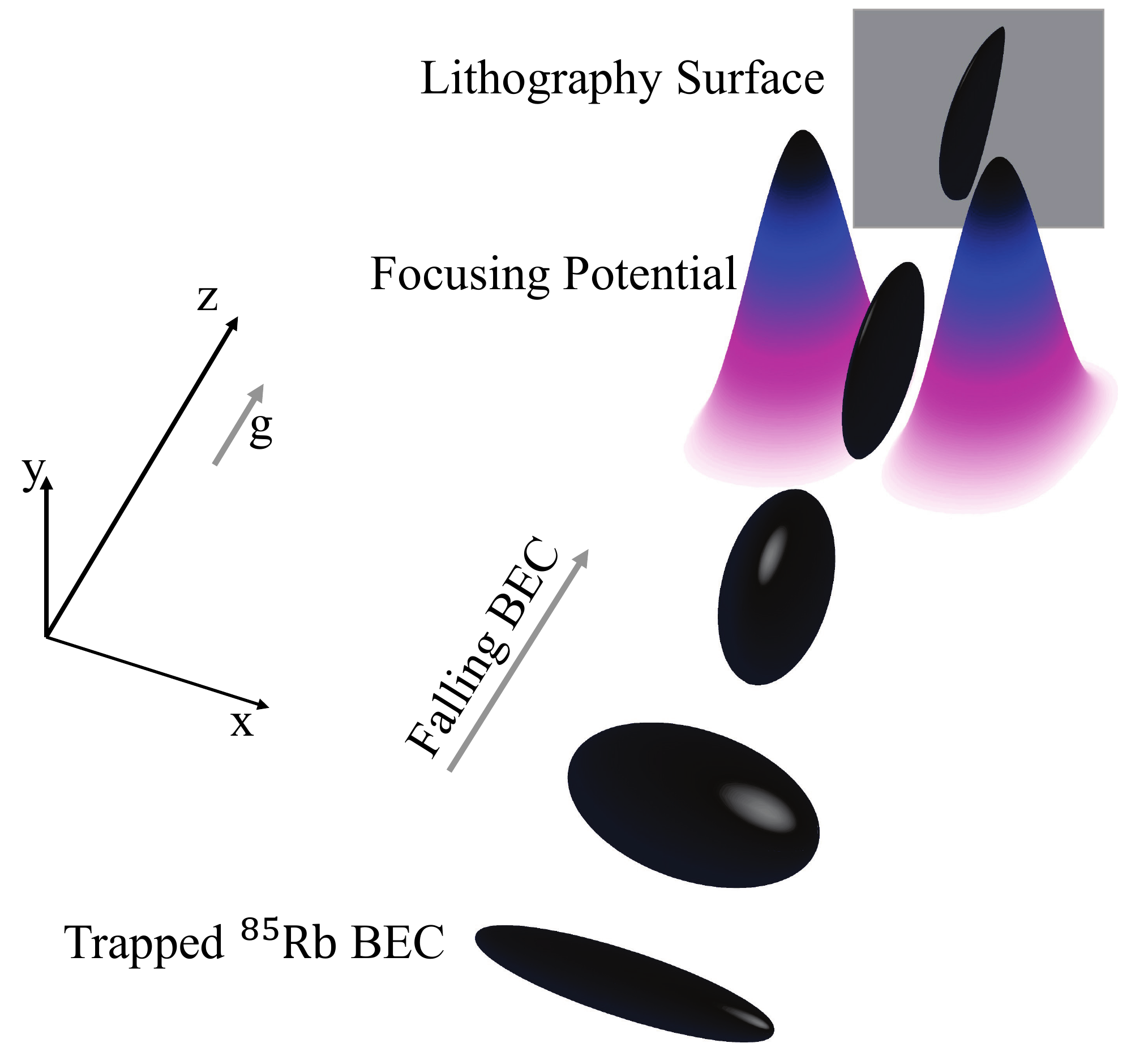}
\caption{3D Schematic illustration of atom deposition using a $^{85}$Rb BEC focused by an optical potential, assumed to be approximately harmonic. The BEC falls along the $z$-axis, which is the direction of gravitational acceleration.}
\label{1}
\end{figure}

The problem we consider is illustrated schematically in Figure~\ref{1}. A cloud of $^{85}$Rb atoms is initially confined by a harmonic trap potential at $t=0$. Having this potential turned off abruptly, the released BEC starts expanding while propagating freely along the vertical $z$ axis while it is approaching the tight harmonic focusing potential along the $x$ axis. It then travels through the potential and is focused down to a nano-meter structure along the $x$ axis, where it is deposited on a surface.

We begin by defining the time-dependent nonlinear Schroedinger equation (i.e. the GPE) \cite{21,22,23,24} in 3D
\begin{equation}
\begin{split}
i\hbar\frac{\partial\psi(\mathbf{r},t)}{\partial t}=\Big(-\frac{\hbar^2}{2m}\nabla_r^2+V_{\text{ext}}(\mathbf{r},t)+u\psi|(\mathbf{r},t)|^{2}\\-iK\psi|(\mathbf{r},t)|^{4}\Big)\psi(\mathbf{r},t),
\end{split}
\label{e1}
\end{equation}
where $\hbar$ and $m$ are Planck's constant and the atomic mass for rubidium-85, respectively. The first nonlinear term, $u|\psi(\mathbf{r},t)|^{2}$, is the mean-field potential term where $u=\frac{4\pi\hbar^2a_s}{m}$ is the inter-atomic interaction strength, $|\psi(\mathbf{r},t)|^{2}$ is the atomic density, and $a_s$ is the $s$-wave scattering length \cite{25, 26, 27, 28}. The value $a_s$ can be tuned in practice from positive (repulsive interactions) to negative values (attractive interactions) through the use of a Feshbach resonance \cite{29}. The second nonlinear term governs three-body recombination losses. In this work we set the three-body loss rate coefficient to $K=4\times 10^{-41}$ m$^6$s$^{-1}$, consistent with experimentally determined values for $^{85}$Rb condensates \cite{34, n6}. At $t=0$, the condensate is initially confined and held by a harmonic trap, $V_{\text{ext}}(\mathbf{r}, t=0)=m(\omega_{0x}^2x^2+\omega_{0y}^2y^2+\omega_{0z}^2z^2)/2$, where $\omega_{0x}$, $\omega_{0y}$, and $\omega_{0z}$ are the initial trapping frequencies along the $x$, $y$ and $z$ axes, respectively, at $t=0$. For $t>0$, the confining potential is switched off and the focusing parabolic potential is switched on. We assume an optical focusing potential, which induces a dipole moment in the atoms of the BEC. The interaction between the dipole moment and the electric field causes a dipole force \cite{37} with a gradient towards the nodes or anti-nodes of the laser intensity. The focusing potential generated by a laser of intensity $I(x, z)$ is \cite{19}
\begin{equation}
U_{\text{dip}}(x,z)=\frac{\hbar\Delta}{2}\ln\Big(1+\frac{\gamma^2}{\gamma^2+4\Delta^2} \frac{I(x,z)}{I_s}\Big),
\label{e2}
\end{equation}
where $\Delta$ denotes the detuning of the laser frequency from the atomic resonance, $\gamma=38$ MHz the natural linewidth of the D$_2$ atomic transition of $^{85}$Rb (i.e. spontaneous decay rate) and $I_s=1.67$ mW/cm$^2$ is the saturation intensity of this transition. An approximately harmonic potential along $x$ can be engineered using a spatial light modulator \cite{38,39}. Assuming a Gaussian beam profile along $z$ (see Figure~\ref{1}), this gives $I(x,z)=I_0 \exp(-2z^2/\sigma_z^2)(k^2x^2)$, where $I_0$ is the maximum intensity of the spatially varying harmonic profile, $\sigma_z$ is the radius of the beam at $1/e^2$ value of the maximum intensity and $k$ determines the strength of the harmonic focusing.

To study the evolving BEC in a focusing potential, we assume that the BEC is located in a stationary frame at $z=0$ while the harmonic potential is in a moving frame approaching the BEC. In this frame, Eq.(\ref{e2}) is
\begin{equation}
U_{\text{dip}}(x,t)=\frac{\hbar\Delta}{2} \ln\Big(1+\frac{\gamma^2}{\gamma^2+4\Delta^2}\frac{I_0}{I_s}k^2x^2f(t)\Big),
\label{e3}
\end{equation}
where
\begin{equation}
f(t)=\exp\Big(\frac{-2}{\sigma_z^2}\big(z_0-z(t)\big)^2\Big),
\label{e4}
\end{equation}
with $z_0$ being the initial distance between the center-of-mass of the condensate and the center of the focusing potential, and $z(t)=\frac{1}{2}gt^2+v_0t$, is the varying distance in terms of time, which depends on the gravitational acceleration, $g$, and initial velocity, $v_0$, imparted to the BEC. For the relatively low values of intensity, $I_0$, and relatively large values of the detuning, $\Delta$, Eq.(\ref{e3}) reduces to  
\begin{equation}
U_{\text{dip}}(x,t)\approx\frac{\hbar\Delta\gamma^2}{(\gamma^2+4\Delta^2)}\frac{I_0}{I_s}k^2x^2f(t).
\label{e5}
\end{equation}
In this regime, Eq.(\ref{e5}) takes the form of an harmonic potential such that $V_{\text{ext}}(x, t>0)=\frac{1}{2}m\omega^2(t)x^2=U_{\text{dip}}(x,t)$, with time-dependent frequency:
\begin{equation}
\omega_x^2(t)=\frac{\hbar\Delta\gamma^2k^2}{m(\gamma^2+4\Delta^2)}\frac{I_0}{I_s}\exp\Big(\frac{-2}{\sigma_z^2}\big(z_0-(\frac{1}{2}gt^2+v_0t)\big)^2\Big).
\label{e6}
\end{equation}

\section{\label{sec:level2}The Variational Approach}

In this section we consider a variational approach based on the GPE to model the BEC dynamics. We adapt the approach considered in Refs \cite{31,32} to account for the effect of three-body losses to the BEC dynamics in focusing regimes.

To begin, we note that the GPE wavefunction that describes our BEC minimizes the action \cite{35}:
\begin{equation}
S=\int\mathcal{L}_{\text{tot}}(\mathbf{r}, t)\ d^3\mathbf{r}\ dt,
\label{e7}
\end{equation}
where $\mathcal{L}_{\text{tot}}$ is the total Lagrangian density \cite{36}
\begin{equation}
\begin{split}
\mathcal{L}_{\text{tot}}(\mathbf{r}, t)=\mathcal{L}+\mathcal{L}_{R}=\frac{i\hbar}{2}\Bigg(\psi^*\frac{\partial\psi}{\partial t}-\psi\frac{\partial\psi^*}{\partial t}\Bigg)-\frac{\hbar^2}{2m}|\mathbf{\nabla}\psi|^2\\-V_{\text{ext}}(\mathbf{r}, t)|\psi|^2-\frac{g}{2}|\psi|^4+\mathcal{L}_R,\quad\quad
\label{e8}
\end{split}
\end{equation}
where $\mathcal{L}_R$ is the Lagrangian density for the three-body recombination term in Eq.(\ref{e1}), defined by
\begin{equation}
\mathcal{L}_{R}=-\frac{1}{3}\big(R\psi^*\big),
\label{e13}
\end{equation}
where
\begin{equation}
R(\psi, \psi^*)=-iK|\psi|^4\psi.
\label{e14}
\end{equation}
In order to minimize the GPE action in Eq.(\ref{e7}), we choose an appropriate single-particle trial wavefunction. A good choice is the following Gaussian variational ansatz
\begin{equation}
\begin{split}
\psi_0(x, y, z, t)=\mathcal{A}(t)\exp\Bigg\{\sum\limits_{r=x, y, z}\quad\quad\quad\quad\quad\quad\quad\quad\quad\quad\quad\quad\quad\quad\\\Bigg[-\frac{\Big(r-r_0(t)\Big)^2}{2\mathcal{W}_r^2(t)}+i\Big(\alpha_r(t)r+\beta_r(t)r^2+\phi_r(t)\Big)\Bigg]\Bigg\},\quad\quad\quad\quad
\label{e9}
\end{split}
\end{equation}
with the normalization factor
\begin{equation}
A(t)=\frac{\sqrt{N(t)}}{\sqrt{\pi^{\frac{3}{2}}\mathcal{W}_x(t)\mathcal{W}_y(t)\mathcal{W}_z(t)}},
\label{e10}
\end{equation}
where $\mathcal{W}_r$, $r_0$, $\beta_r$, $\alpha_r$ and $\phi_r$ are the variational parameters. $\mathcal{W}_r$ corresponds to the respective condensate width, $r_0$ is the initial position of the BEC center, $\beta_r$ indicates the (curvature radius$)^{-1/2}$, $\alpha_r$ represents the slope, and $\phi_r$ describes the phase of the condensate. The term $N(t)$ represents the number of atoms in the BEC, which changes with time. This follows from $\int d^3\mathbf{r}\ |\psi_0(\mathbf{r},t)|^2$ which is not necessarily conserved due to three-body losses.  We set the center of the condensate to the center of the Cartesian coordinates, $(x_0, y_0, z_0)=(0, 0, 0)$.

The objective is to find the equations of motion for these parameters. Substituting Eq. (\ref{e8}) and Eq. (\ref{e9}) into the average Lagrangian density, $L_{\text{tot}}(t)=\int d^3\mathbf{r}\ \mathcal{L}_{\text{tot}}(\mathbf{r}, t)$, we obtain (see Appendix 1 for details)
\begin{equation}
\small{\begin{split}
L_{\text{tot}}(t)=-\sum\limits_{r=x, y, z}\frac{|A|^2\mathcal{W}_r\sqrt{\pi}}{2}\Bigg\{\hbar\dot{\beta}_r\mathcal{W}_r^2+2\hbar\dot{\phi}_r+\frac{\hbar^2}{2m}\big(\frac{1}{\mathcal{W}_r^2}\big)\\+\frac{\hbar^2}{2m}\big(4\beta_r^2\mathcal{W}_r^2\big)+\frac{\hbar^2}{2m}\big(2\alpha_r^2\big)+\frac{1}{2}m\omega_x^2(t)\mathcal{W}_x^2+\frac{g|A|^2}{\sqrt{2}}\Bigg\}+L_R(t),
\label{e12}
\end{split}}
\end{equation}
where
$L_R(t)=\int d^3\mathbf{r}\ \mathcal{L}_{R}(\mathbf{r}, t)$. 
The equations of motion for all variational factors are given by the Euler-Lagrange equations for real and imaginary components of $L$. Since the imaginary part is trivial, one can ignore it and only consider the contribution due to the real components
\begin{equation}
\frac{d}{dt}\frac{\partial L}{\partial\dot{q_i}}-\Big[\frac{\partial L}{\partial q_i}+\Re\bigg(\frac{\partial L_R}{\partial q_i}\bigg)\Big]=0;\ q\in\{\mathcal{A}(t), \mathcal{W}_r, \alpha_r, \beta_r, \phi_r\},
\label{e15}
\end{equation}
where $\frac{\partial L_{\text{tot}}}{\partial\dot{q_i}}=\frac{\partial L}{\partial\dot{q_i}}$ since $\frac{\partial L_R}{\partial\dot{q_i}}=0$.
Since $\frac{\delta\mathcal{L}_{R}}{\delta\psi^*}=-\frac{1}{3}R$, we can write
 \begin{equation}
\Re\Big(\frac{\partial\mathcal{L}_{R}}{\partial q_i}\Big)=-\frac{1}{6}\Big(R\frac{\partial\psi^*}{\partial q_i}+R^*\frac{\partial\psi}{\partial q_i}\Big).
\label{e16}
\end{equation}
Integrating both sides of Eq.(\ref{e16}) and inserting the associated result into Eq.(\ref{e15}) gives 
\begin{equation}
\frac{d}{dt}\frac{\partial L}{\partial\dot{q_i}}-\frac{\partial L}{\partial q_i}=-\frac{1}{6}\int \Big(R\frac{\partial\psi^*}{\partial q_i}+R^*\frac{\partial\psi}{\partial q_i}\Big)\ d^3\mathbf{r}.
\label{e17}
\end{equation}
Solving Eq.(\ref{e17}) for $q_i=\mathcal{A}, \mathcal{W}_r, \alpha_r, \beta_r, \phi_r$ results in the following dimensionless variational equations for the BEC width dynamics and loss rate (see Appendix 2 for details)
\begin{equation}
	\begin{split}
		\frac{d^2{\mathcal{W}}_x}{dt^2}+\omega_x^2(t)\mathcal{W}_x=\frac{\hbar^2}{m^2}\Big(\frac{1}{\mathcal{W}_x^3}\Big)+\frac{gN}{m(2\pi)^{3/2}\mathcal{W}_x^2\mathcal{W}_y\mathcal{W}_z}\\
		-\frac{7K^2N^4}{3(3\pi)^6\hbar^2\mathcal{W}_x^3\mathcal{W}_y^4\mathcal{W}_z^4}.~~~~~~~~~~~~~~~~~
		\label{e18}
	\end{split}
\end{equation}
\begin{equation}
	\begin{split}
		\frac{d^2{\mathcal{W}}_y}{dt^2}=\frac{\hbar^2}{m^2}\Big(\frac{1}{\mathcal{W}_y^3}\Big)+\frac{gN}{m(2\pi)^{3/2}\mathcal{W}_x\mathcal{W}_y^2\mathcal{W}_z}\\
		-\frac{7K^2N^4}{3(3\pi)^6\hbar^2\mathcal{W}_x^4\mathcal{W}_y^3\mathcal{W}_z^4},~~~~~~~~~~~~~~~~~
		\label{e19}
	\end{split}
\end{equation}
\begin{equation}
	\begin{split}
		\frac{d^2{\mathcal{W}}_z}{dt^2}=\frac{\hbar^2}{m^2}\Big(\frac{1}{\mathcal{W}_z^3}\Big)+\frac{gN}{m(2\pi)^{3/2}\mathcal{W}_x\mathcal{W}_y\mathcal{W}_z^2}\\
		-\frac{7K^2N^4}{3(3\pi)^6\hbar^2\mathcal{W}_x^4\mathcal{W}_y^4\mathcal{W}_z^3},~~~~~~~~~~~~~~~~~
		\label{e20}
	\end{split}
\end{equation}
\begin{equation}
	\frac{dN(t)}{dt}=-\frac{KN^3}{9\sqrt{3}\pi^3\hbar\mathcal{W}_x^2\mathcal{W}_y^2\mathcal{W}_z^2},
	\label{e21}
\end{equation}
with the initial conditions
\begin{equation}
	\begin{split}
		\mathcal{W}_x(t=0)=\mathcal{W}_{0x};\ \mathcal{W}_y(t=0)=\mathcal{W}_{0y};\ \mathcal{W}_z(t=0)=\mathcal{W}_{0z};\\
		\dot{\mathcal{W}}_x(t=0)=\dot{\mathcal{W}}_{0x};\ \dot{\mathcal{W}}_y(t=0)=\dot{\mathcal{W}}_{0y};\ \dot{\mathcal{W}}_z(t=0)=\dot{\mathcal{W}}_{0z},\\
		N(t=0)=N_0.\quad\quad\quad\quad\quad\quad\quad\quad\quad
		\label{e22}
	\end{split}
\end{equation}
 We estimate the initial BEC widths along all three Cartesian axes, $\mathcal{W}_{0x},\  \mathcal{W}_{0y}, \ \mathcal{W}_{0z}$, using \cite{27}
\begin{equation}
\mathcal{W}_{0i}=\Bigg(\frac{2}{\pi}\Bigg)^{1/10}\Bigg(\frac{N_0a_s}{l}\Bigg)^{1/5}\frac{\omega_0}{\omega_i}\ l,\quad\quad (i=x, y, z)
\label{e25}
\end{equation}
in which we have introduced the harmonic oscillator length $l=\sqrt{\frac{\hbar}{m\omega_{0}}}$, where $\omega_0=(\omega_{0x}\omega_{0y}\omega_{0z})^{1/3}$. This estimate is obtained by minimizing the energy of a 3D Gaussian ansatz and neglecting the kinetic energy contribution, which is small for strong repulsive initial interactions.

\begin{figure*}[tb]
	\centering
	\includegraphics[width=19cm, height=7.5cm,angle=0]{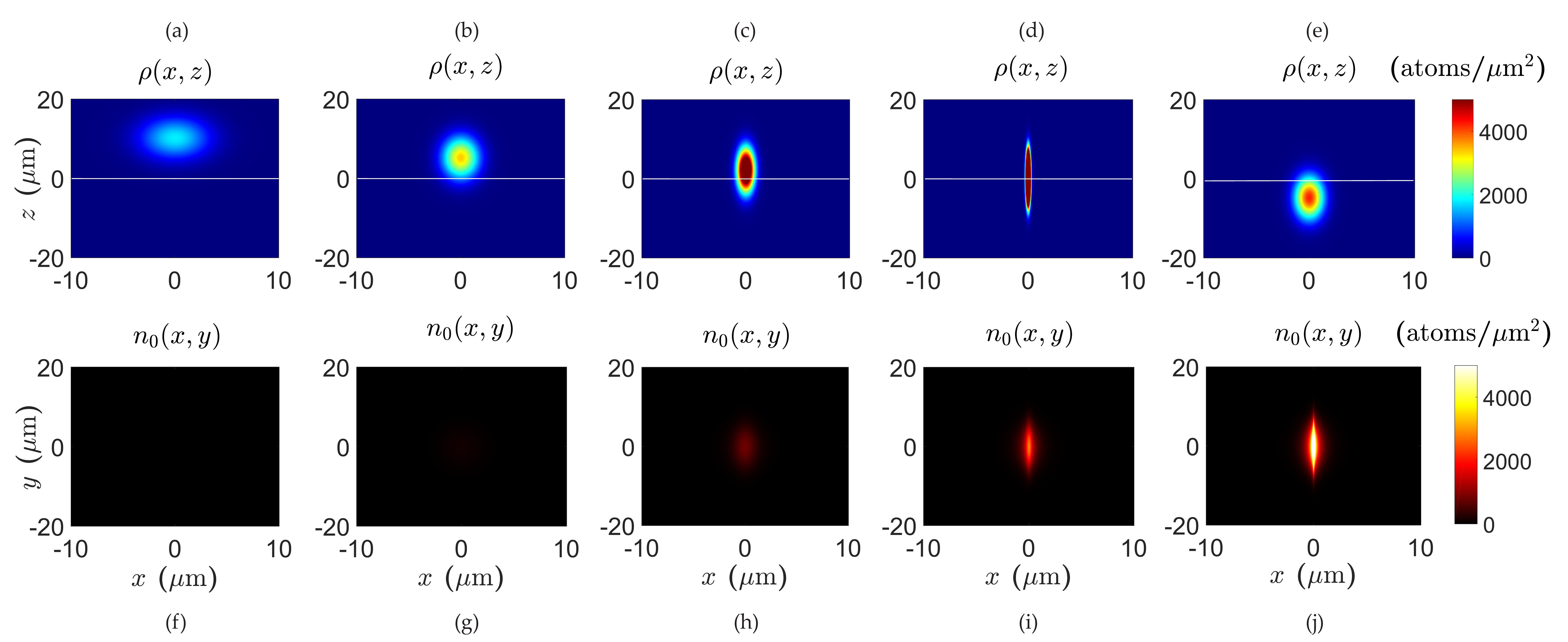}
	\caption{Upper row: The cross section view of the evolving BEC column density $\rho(x, z)=\int dy\rho(x, y, z)$ in the plane of $(x, z)$. (a)-(d) indicate the procedure of focusing when the center of the BEC is located at $z_0=10, 5, 2, 0$ and $-5~\mu$m respectively (from left to right). Lower row: The corresponding integrated density distributions $n_0(x, y)$ at the focal surface $z=0$ (where the substrate is essentially placed). The value of $n_0(x, y)$ reaches to its maximum when all atoms of the cloud have been deposited on the focal surface. The momentum kick in the simulations is set to zero and $a_s=-1a_0$. Parameters are: $N_0=10^5$, $\omega_{0x}=2\pi\times 10$ Hz, $\omega_{0y}=\omega_{0z}=2\pi\times 70$ Hz, $p=32\hbar k$, $\sigma_z=100~\mu$m and $k = 2.01384 \times 10^4$ m$^{-1}$.}
	\label{3}
\end{figure*}

\begin{figure}
	\centering
	\hskip -4ex
	\includegraphics[width=8.5cm, height=13.5cm,angle=0]{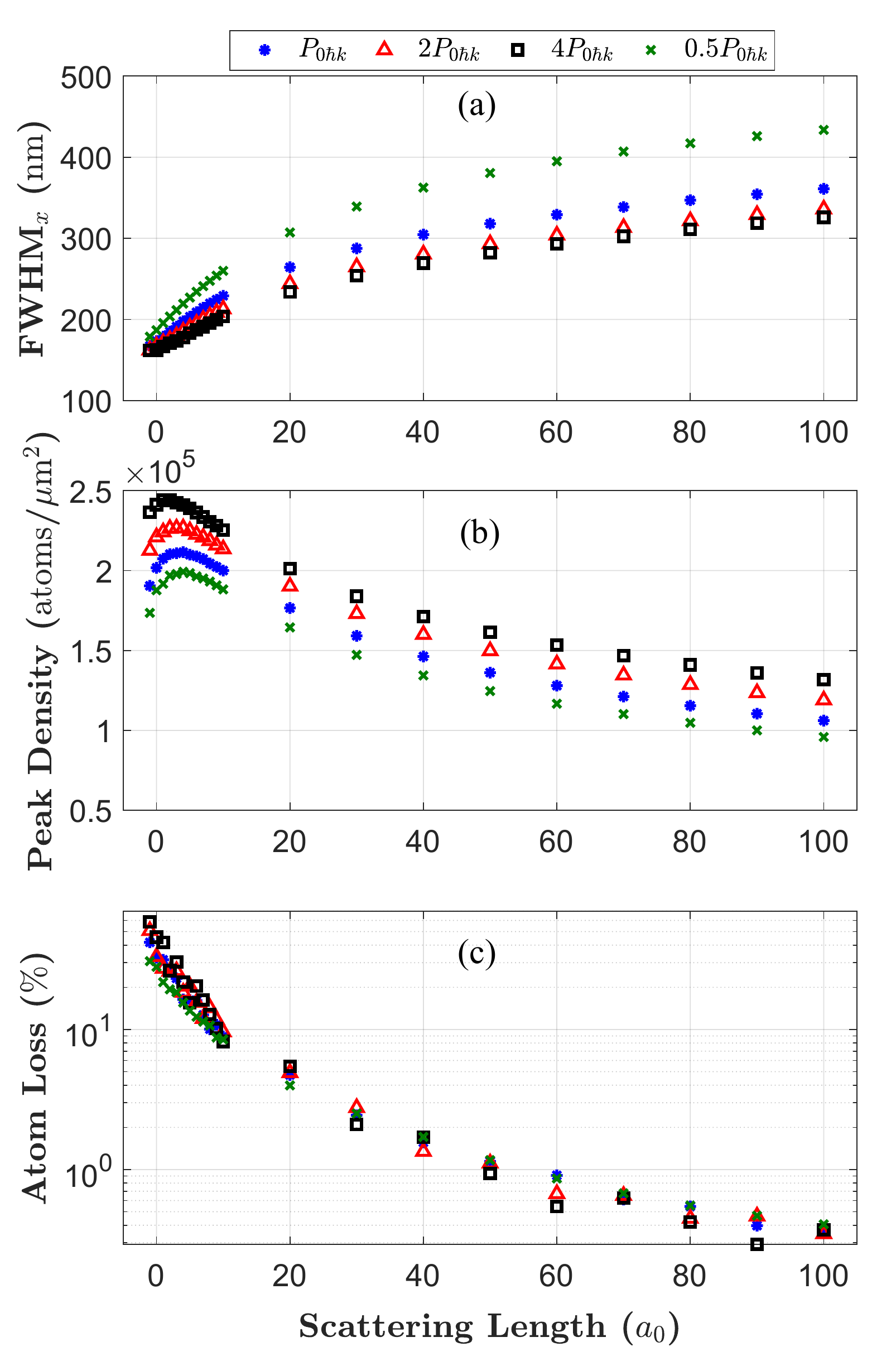}
	\caption{Characterisation of deposited 2D BEC density at different focal planes, determined from variational analysis for different values of scattering length and focusing power. The values of FWHM, integrated peak densities, and atom loss due to three-body losses are shown in (a), (b) and (c) respectively. The focal plane for $0.5P_{0\hbar k}$, $P_{0\hbar k}$, $2P_{0\hbar k}$ and $4P_{0\hbar k}$ is, respectively, located at $z_f=-31.3$, $0$, $23.9$ and $44.3~\mu$m along the $z$ axis.
		Parameters used are: $N_0=10^5$, $\sigma_z=100~\mu$m, $k = 2.01384 \times 10^4$ m$^{-1}$, $\Delta=200$ GHz, $I_s=16.7$ W/m$^2$, $\gamma=38$ MHz, $v_0=0$, $v(z=0)=9.9$ cm/s, $P_{0\hbar k}= 4.127$ mW, $\mathcal{W}_{0x}=19.6~\mu$m, $\mathcal{W}_{0z}=\mathcal{W}_{0y}=2.8~\mu$m, $\dot{\mathcal{W}}_{0z}=\dot{\mathcal{W}}_{0z}=\dot{\mathcal{W}}_{0z}=0$, $a_0=5.29\times 10^{-11}$ m and $K=4\times 10^{-41}$m$^6$s$^{-1}$.}
	\label{4}
\end{figure}

\begin{figure}[ht!]
	\vskip 1ex
	\hskip -6ex
	\includegraphics[width=9.5cm, height=6.8cm,angle=0]{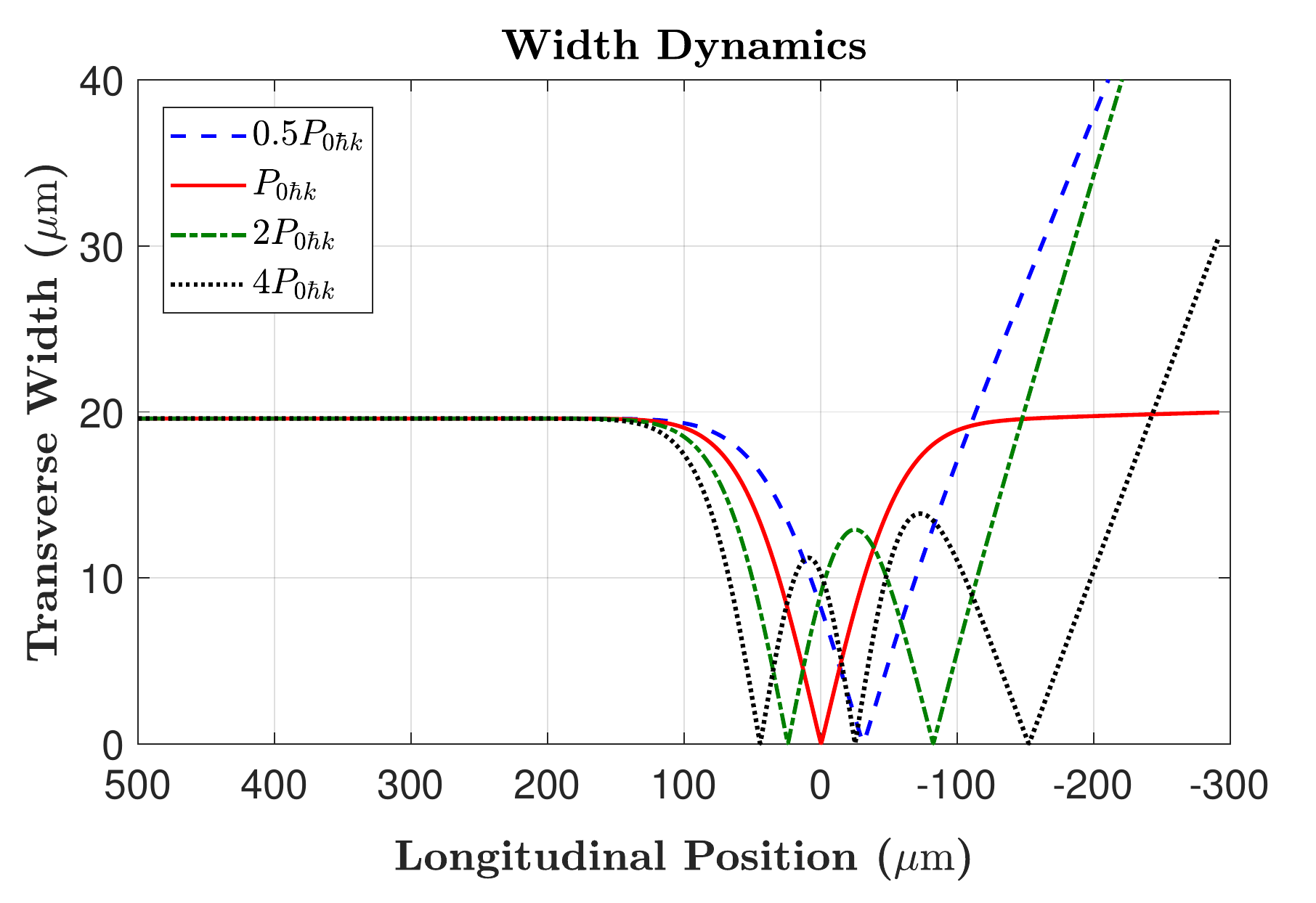}
	\caption{Transverse width dynamics as a function of the longitudinal direction ($z$ axis) for a focusing BEC dropped at $z_0=500~\mu$m with zero momentum kick, $p=0\hbar k$. The blue dashed, red solid, green dash-dotted and black dotted curves, respectively, correspond to $0.5P_{0\hbar k}$, $P_{0\hbar k}$, $2P_{0\hbar k}$ and $4P_{0\hbar k}$ where $P_{0\hbar k}$ is the optimal power to focus the BEC at $z=0$, and is estimated as $4.127$ mW. The lowest minimum peak (the focal plane) for $0.5P_{0\hbar k}$, $P_{0\hbar k}$, $2P_{0\hbar k}$ and $4P_{0\hbar k}$ is, respectively, $z_f=-31.3$, $0$, $23.9$ and $44.3~\mu$m.
		Parameters are: $N_0=10^5$, $\sigma_z=100~\mu$m, $k = 2.01384 \times 10^4$ m$^{-1}$, $\Delta=200$ GHz, $I_s=16.7$ W/m$^2$, $\gamma=38$ MHz, $\mathcal{W}_{0x}=19.6~\mu$m, $\mathcal{W}_{0z}=\mathcal{W}_{0y}=2.8~\mu$m, $a_s=100a_0$, $a_0=5.29\times 10^{-11}$ m and $K=4\times 10^{-41}$m$^6$s$^{-1}$.}
	\label{44}
\end{figure}

\begin{figure*}[t!]
	\centering
	\includegraphics[width=18cm, height=9cm,angle=0]{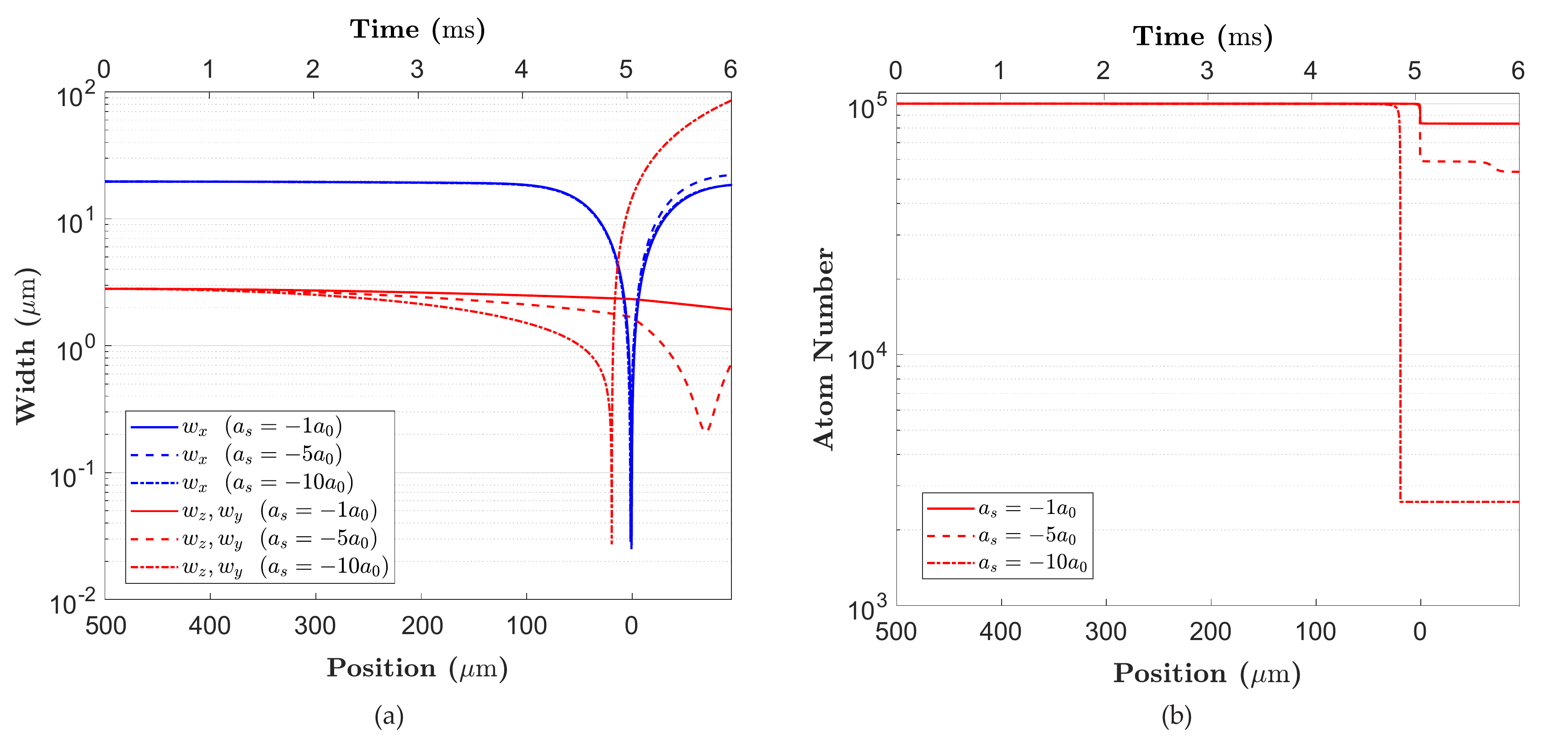}
	\caption{(a): The width dynamics for $a_s= -10a_0$ (dash-dotted curves), $-5a_0$ (dashed curves), and $-1a_0$ (solid curves), for a BEC optimally focused at $z=z_f=0$. The blue curves illustrate the BEC width evolution along the $x$ axis wheras the red curves represent this trend along the $y$ and $z$ axes. (b): The corresponding atom number evolution for the same three scattering lengths.
		Parameters are: $N_0=10^5$, $z_0=500~\mu$m, $\sigma_z=100~\mu$m, $k = 2.01384 \times 10^4$ m$^{-1}$, $\Delta=200$ GHz, $\mathcal{W}_{0x}=19.6~\mu$m, $\mathcal{W}_{0z}=\mathcal{W}_{0y}=2.8~\mu$m, $a_0=5.29\times 10^{-11}$ m and $K=4\times 10^{-41}$m$^6$s$^{-1}$.}
	\label{5}
\end{figure*}

\begin{figure}[ht!]
	\centering
	\hskip -10ex
	\vskip -2ex
	\includegraphics[width=9.5cm, height=14cm,angle=0]{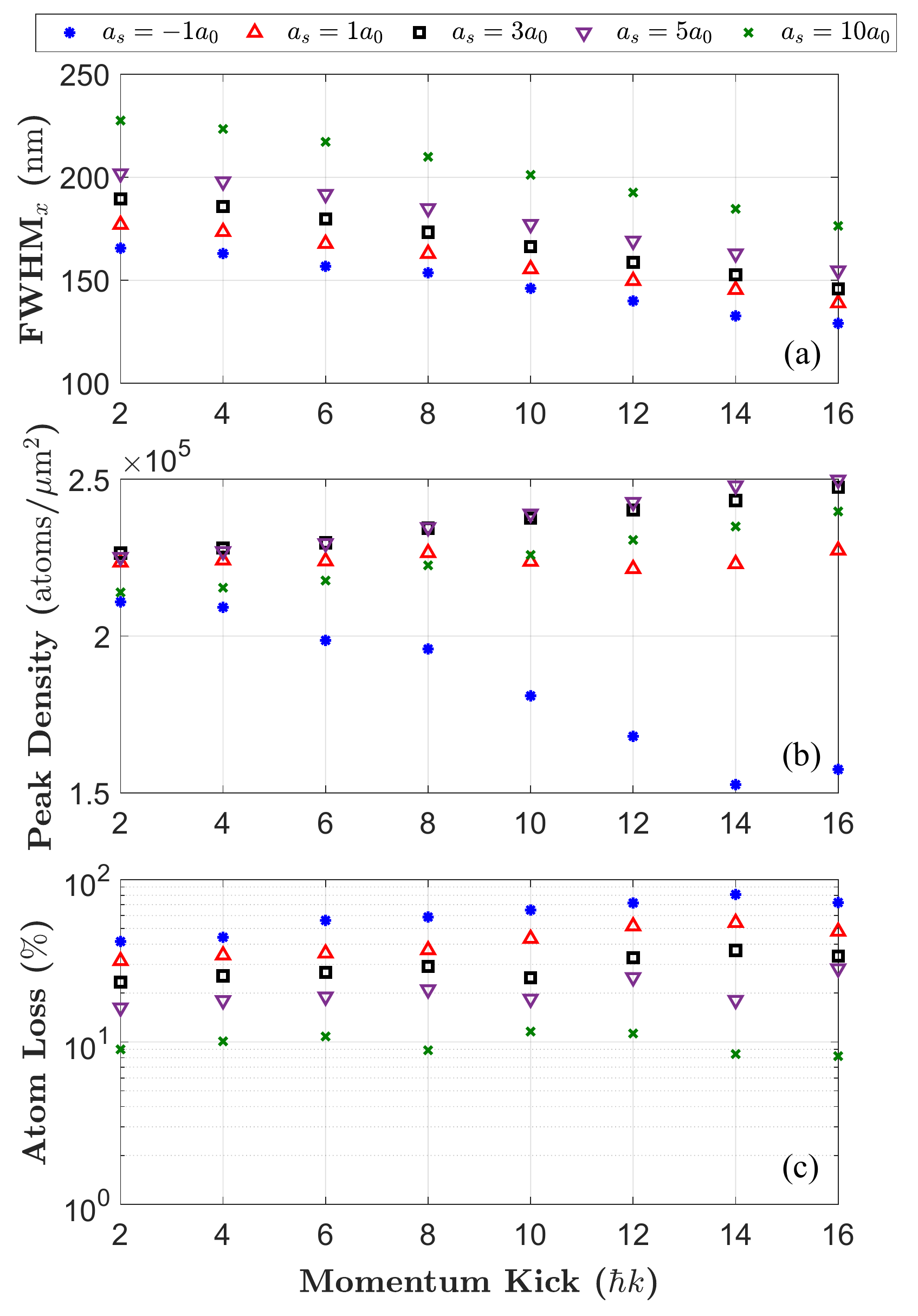}
	\caption{Characterisation of deposited 2D BEC density at $z=0$, determined from the variational analysis for different of momentum kicks and scattering lengths. The values of FWHM, integrated peak density, and the amount of atom loss are plotted in (a), (b) and (c) respectively. 
		Parameters are: $N_0=10^5$, $z_0=500~\mu$m, $\sigma_z=100~\mu$m, $k = 2.01384 \times 10^4$ m$^{-1}$, $\Delta=200$ GHz, $I_s=16.7$ W/m$^2$, $\gamma=38$ MHz , $\mathcal{W}_{0x}=19.6~\mu$m, $\mathcal{W}_{0z}=\mathcal{W}_{0y}=2.8~\mu$m, $\dot{\mathcal{W}}_{0z}=\dot{\mathcal{W}}_{0z}=\dot{\mathcal{W}}_{0z}=0$, $a_0=5.29\times 10^{-11}$ m and $K=4\times 10^{-41}$m$^6$s$^{-1}$ for Rb-85.}
	\label{6}
\end{figure}

\begin{figure}[ht!]
	\hskip -2ex
	\vskip -2ex
	\includegraphics[width=9.5cm, height=14cm,angle=0]{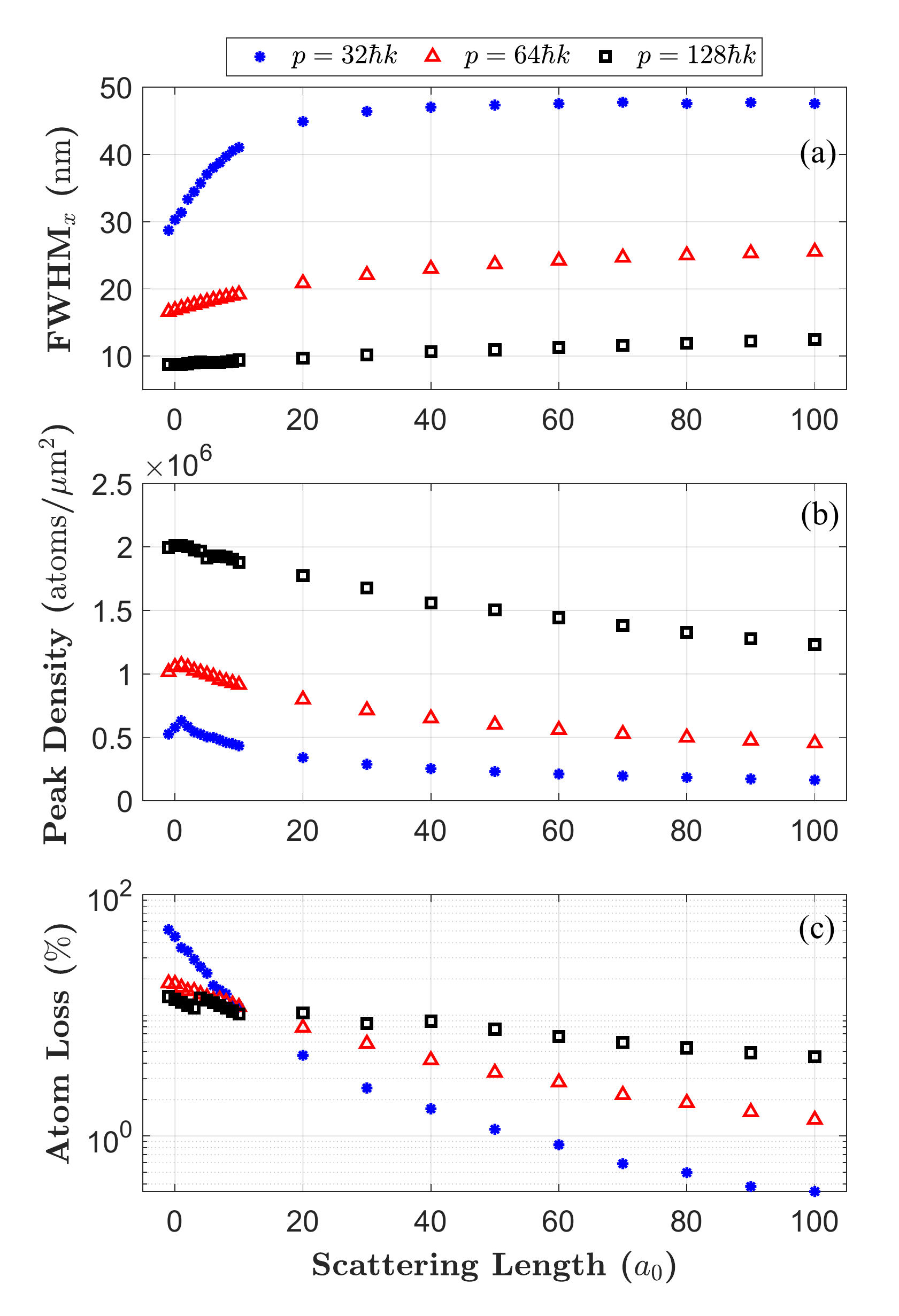}
	\caption{The simulation results of the instantaneous focused profile at $z=0$ when no substrate is considered. The results are determined from the variational method for different momentum kicks and scattering lengths. The values of FWHM, integrated peak density, and the amount of atom loss are calculated in (a), (b) and (c) respectively. Parameters are: $N_0=10^5$, $z_0=500~\mu$m, $\sigma_z=100~\mu$m, $k = 2.01384 \times 10^4$ m$^{-1}$, $\Delta=200$ GHz, $I_s=16.7$ W/m$^2$, $\gamma=38$ MHz , $\mathcal{W}_{0x}=19.6~\mu$m, $\mathcal{W}_{0z}=\mathcal{W}_{0y}=2.8~\mu$m, $\dot{\mathcal{W}}_{0z}=\dot{\mathcal{W}}_{0z}=\dot{\mathcal{W}}_{0z}=0$, $a_0=5.29\times 10^{-11}$ m and $K=4\times 10^{-41}$m$^6$s$^{-1}$ for Rb-85.}
	\label{7}
\end{figure}

\begin{figure*}[ht!]
	\centering
	\includegraphics[width=16cm, height=6.5cm,angle=0]{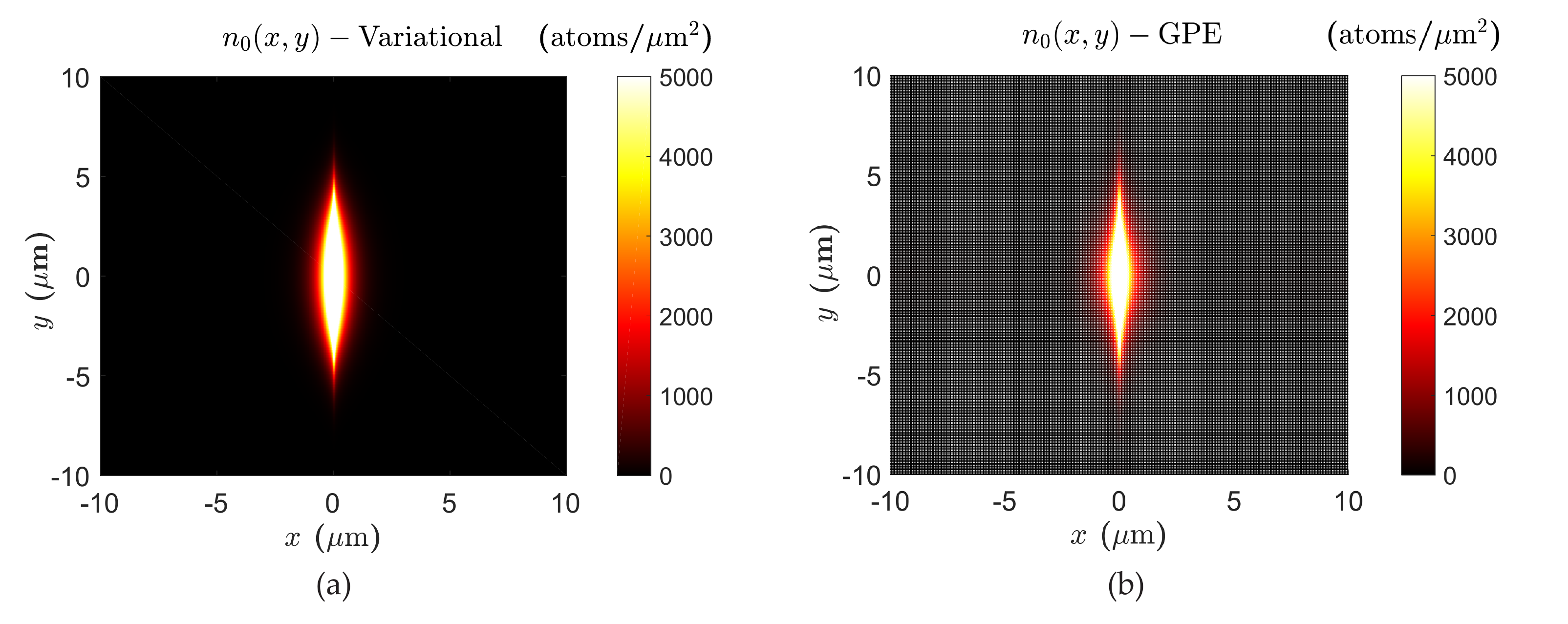}
	\caption{(a): Deposited BEC focused structure on the ($x, y$) plane ($z=0$) carried out via the variational approach. (b): Deposited BEC profile on the same plane acquired by a numerical GPE simulation. In both simulations no momentum kick is applied to the BEC and $a_s=-1a_0$. 
		Parameters are: $N_0=10^5$, $\omega_{0x}=2\pi\times 10$ Hz, $\omega_{0y}=\omega_{0z}=2\pi\times 70$ Hz, $p=32\hbar k$, $\sigma_z=100~\mu$m and $k = 2.01384 \times 10^4$ m$^{-1}$.}
	\label{8}
\end{figure*}

\begin{figure}[t!]
	\centering
	\includegraphics[width=9cm, height=14cm,angle=0]{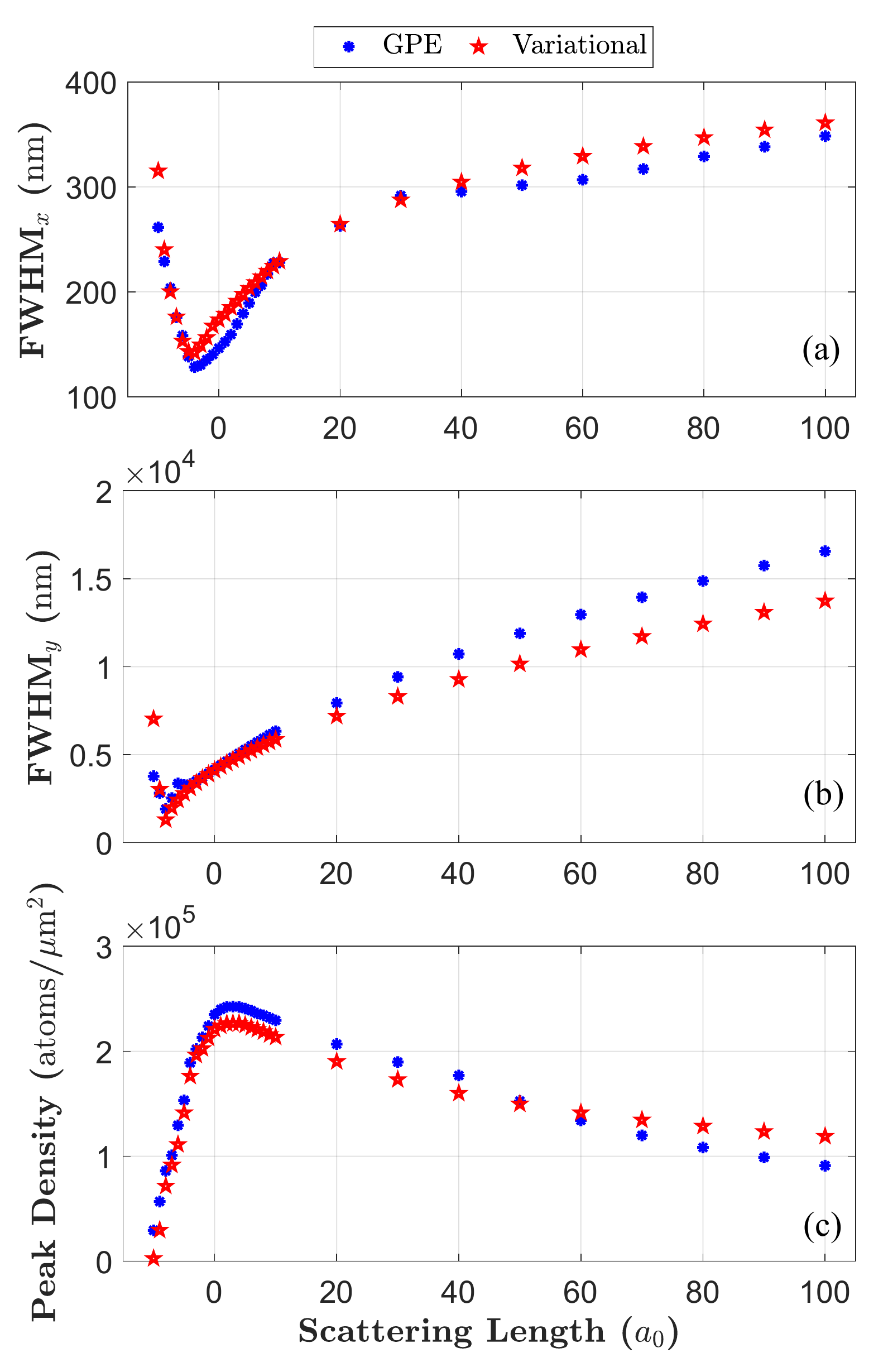}
	\caption{(a): The values of FWHM of the focused distribution along the $x$ axis for different $a_s$. (b): The FWHMs along one of the radial axes, $y$, against various scattering lengths. (c): The achieved values of peak densities of the focused profile when using different interaction strengths. The red dots in all three graphs indicate predictions from the variational method, whereas the blue dots show results from the numerical GPE simulations.
		Parameters are: $N_0=10^5$, $z_0=500~\mu$m, $\sigma_z=100~\mu$m, $k = 2.01384 \times 10^4$ m$^{-1}$, $\Delta=200$ GHz, $I_s=16.7$ W/m$^2$, $\gamma=38$ MHz, $p=0\hbar k$, $P_{0\hbar k}= 4.127$ mW, $a_0=5.29\times 10^{-11}$ m and $K=4\times 10^{-41}$m$^6$s$^{-1}$.}
	\label{9}
\end{figure}

\begin{figure}[t!]
	\centering
	\includegraphics[width=9.5cm, height=8.5cm,angle=0]{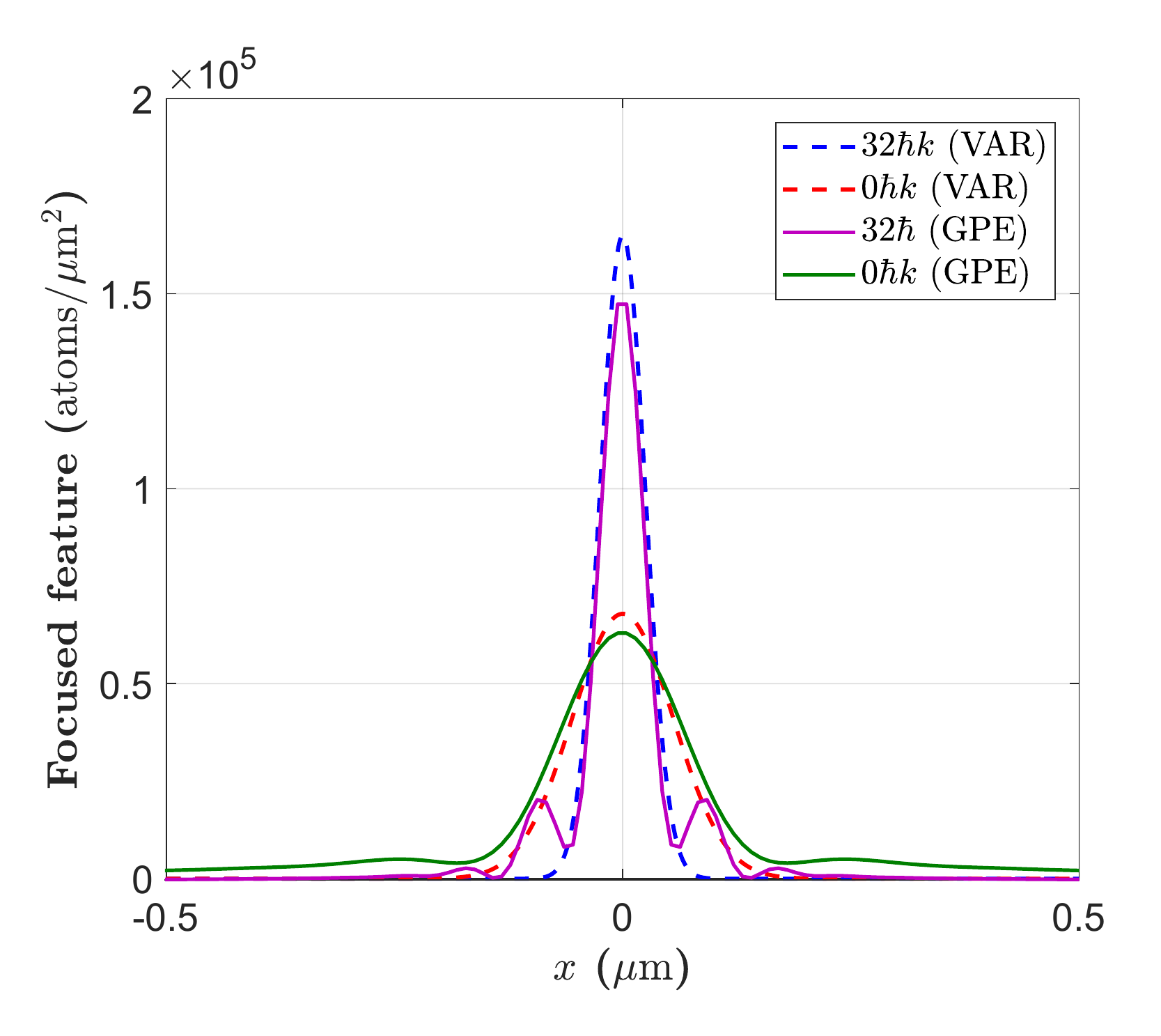}
	\caption{BEC focused structures, $n_0(x)$, along the transverse $x$ axis on the planes $y=0$ and $z=0$ achieved via the variational approach (dashed curves) and GPE simulations (solid curves) for $p=0$ (green and red curves) and $32\hbar k$ (blue and purple curves). The BEC scattering length is set to $100a_0$ in all the simulations.}
	\label{10}
\end{figure}

\section{\label{sec:level3}Optimal Power for Focusing}

By treating the atom dynamics as classical particle trajectories, the laser power needed to focus the atoms at any desired spot along the focal axis ($z$-axis) can be determined \cite{20}
\begin{equation}
P_0=\xi\frac{\pi}{4}\frac{E_0}{\hbar\Delta}\frac{\gamma^2+4\Delta^2}{\gamma^2}\frac{I_s}{k^2},
\label{e29}
\end{equation}
where $E_0$ is the initial kinetic energy of the atoms and $\xi$ is a dimensionless parameter. The power relates to the peak intensity via $I_0=8P_0/\pi\sigma_z^2$. A value of $\xi=5.37$, determined by solving the classical equations of motion for atomic trajectories, optimally focuses the atoms onto the plane $z=0$ and $x=0$ (the center of the focusing potential). The selection of lower values of $\xi$ (smaller powers) leads to focusing on planes $z< 0$.

\section{\label{sec:level4}Estimate of deposited 2D density distribution}

Suppose we have a substrate placed at the $z = 0$ plane. We consider an ideal atom lithographic scenario, where atoms that intersect the $z = 0$ plane are deemed deposited on the substrate surface. A time integral of the density profile at $z = 0$ therefore provides a simple estimate of the total 2D distribution of atoms deposited on this surface:
\begin{equation}
n_0(x,y)=\int_{0}^{t_{\text{end}}}|\psi(x, y, t, z=0)|^2\ dt,
\label{e30}
\end{equation} 
where $t_\text{end}$ is the duration of the atom lithographic process. This estimate neglects the effect of the surface itself on the BEC dynamics. For example, as atoms are deposited on the surface, they leave the condensate and lower the overall mean-field energy. Furthermore, atoms that are not deposited will be reflected, potentially impacting the focussing dynamics. Nevertheless, the estimate of the deposited 2D density distribution provided by Eq.(\ref{e30}) can be considered a ``best case''.

Our goal is to determine atomic and focusing potential parameters that result in narrow deposited distributions with a high peak density. These can be estimated from Eq. (26) via the full width at half maximum (FWHM) of $n_0(x,y)$ and $\max_{x,y} n_0(x,y)$, respectively. Throughout this paper, we choose $\xi = 5.37$ since this provides optimal focusing in the $z=0$ plane.

The principle of BEC deposition is illustrated in Figures~\ref{3}(a)-(j). In the top row, the density profile of the released BEC in the 2D $(x, z)$ plane is shown (the profile has been integrated over the $y$ axis). As it falls through the optical potential, it becomes more focused along the $x$ axis until it reaches its focal spot at $z=0$ [see  Figures~\ref{3}(a)-(d)]. It then begins to expand once it leaves the focal plane, $z=0$ [see  Figure~\ref{3}(e)].  The second row [Figures~\ref{3}(f)-(j)] shows the time integrated density in the $z=0$ plane, which we interpret as the accumulated or deposited density on the surface. As seen from left to right, this increases until all atoms have been deposited on the surface.

\section{\label{sec:level5}Results}

We assume an initial cylindrical BEC of $N_0=10^5$ $^{85}$Rb atoms confined by a harmonic trap of axial frequency $\omega_{0x}=2\pi\times 10$ Hz and radial frequency  $\omega_{0r}=\omega_{0z}=\omega_{0y}=2\pi\times 70$ Hz. The center of the trap is located at $z_0=500~\mu$m from the center of the focusing harmonic potential. $^{85}$Rb has an easily tuneable Feshbach resonance \cite{n7}. We initially set the $s$-wave scattering to $a_s= 100 a_0$ ($a_0$ is the Bohr radius), giving a condensate with large mean-field energy and an inverted-parabolic Thomas-Fermi density profile. Once the trap is switched off, and the BEC is allowed to freely propagate towards the focusing potential, we quench the scattering length to a value between $a_s= -10 a_0$ (attractive) and $a_s= 100 a_0$ (repulsive). In the simplest case, the atoms are simply dropped from the trap and fall under gravity. We also consider scenarios where a momentum kick is imparted to the atoms upon release, which could be achieved with either a Raman or Bragg \cite{n5} optical transition. All results presented in this paper used a focusing potential with $\sigma_z = 100~\mu$m, $k = 2.01384 \times 10^4$ m$^{-1}$, and $\Delta = 200$ GHz, and use a 3-body recombination loss rate coefficient of $K = 4 \times 10^{-41}$ m$^6$ s$^{-1}$ \cite{34}.

Figure~\ref{4} shows the results of the integrated peak densities, FWHM, and atom number variation for different two-body interaction strengths, from $a_s=-1a_0$ to $a_s=100a_0$, when there are no momentum kicks. In this case, the required optimal power to focus the BEC at $z=0$ (center of the potential, the focal plane) is estimated as $P_{0\hbar k}= 4.127$ mW by Eq.(\ref{e29}). For a comparison, results are also reproduced for different powers. We note that changing the field power for the same momentum kick (i.e. $p=0\hbar k$) causes the focal plane to be shifted along the $z$ axis such that for $P>P_{0\hbar k}$ and $P<P_{0\hbar k}$, the focal plane is located, respectively, above and below the $z=0$ plane. Figure~\ref{44} represents the width dynamics of the BEC along the $x$ (transverse) axis as a function of the longitudinal position for four various powers, $0.5P_{0\hbar k}$, $P_{0\hbar k}$, $2P_{0\hbar k}$ and $4P_{0\hbar k}$. Notice that for $P>P_{0\hbar k}$ a breathing-like oscillation of the BEC occurs as it passes through the focusing potential along the $z$ axis. In such an event, multiple minimum peaks appear in the width dynamics. For $P=2P_{0\hbar k}$ (green dash-dotted curve) and $4P_{0\hbar k}$ (black dotted curve) the lowest minimum is $z_f=23.9~\mu$m and $z_f=44.3~\mu$m, respectively, occurring above the plane of $z=0$. For $P\leqslant P_{0\hbar k}$, however, there always exists a single minimum peak, which is located at $z_f=0$ (red solid curve) and $-31.3~\mu$m (blue dashed curve) for $P_{0\hbar k}$ and $0.5P_{0\hbar k}$ respectively. Hence, varying the power allows us to compare the quality of the deposited 2D BEC density at different focal planes; results of this comparison are shown in Fig.~\ref{4}.

As illustrated in Figure~\ref{4}(a), for a fixed power value, a decrease in the scattering length reduces the FWHM along the focusing $x$ direction. In other words, narrower and finer structures are given by smaller \textit{s}-wave interactions. This variation becomes more significant as the BEC interactions switch from repulsive to attractive. For $a_s>5a_0$, decreasing the scattering length results in higher peak densities, since repulsive interactions hinder the focusing of the atoms. However, for $a_s\leqslant 5a_0$, lowering $a_s$ causes the peak densities to decrease, see Figure~\ref{4}(b). This is due to the three-body recombination losses which become increasingly important as the density of the condensate increases. This explains the loss rate of atoms in Figure~\ref{4}(c), which increases as the \textit{s}-wave scattering length is reduced. Although three-body losses minimally affect a BEC with large repulsive interactions, they become significant in higher density regimes for $a_s\leqslant 10a_0$, where up to 60$\%$ of the atoms can be lost. Altering the laser power also substantially affects the deposition. According to Figure~\ref{4}(a-c), increasing the power leads to a better resolution (smaller FWHM) and larger peak density for each value of $a_s$, at the expense of moving the focal plane.

We now investigate in more detail the effect of attractive interactions for $-10a_0\leqslant a_s\leqslant -1a_0$. In this regime, small decreases in the value of the scattering length can considerably alter the density profile. Fig~\ref{5}(a) compares the BEC widths, along all  $x$, $y$ and $z$ axes for different scattering lengths $a_s=-10, -5, -1 a_0$ as a function of longitudinal position, $z$. In this figure the BEC is prepared at $z_0=500~\mu$m and dropped (no momentum kick) at $t=0$. The minimum possible value for $\mathcal{W}_x$ for all three interaction strengths almost occurs at $z=0$ (see the overlap between the blue solid, dashed and dash-dotted curves). However, the resultant radial widths, $\mathcal{W}_y$ and $\mathcal{W}_z$, are quite different from one scattering length to another (see the red solid, dashed and dash-dotted curves). As revealed, the minimum peak in the $\mathcal{W}_y$ and $\mathcal{W}_z$ curves is shifted closer to $z=0$ as $a_s$ is reduced such that for $a_s=-10a_0$, this point takes place before $z=0$ ($z<0$). This effect is well explained by the collapse of a BEC in high density regimes \cite{58, 59, n6} where the attractive interactions between atoms cause the BEC to rapidly and strongly shrink at a critical density and then to expand sharply. As shown in  Fig~\ref{5}(b), as the BEC collapses there is consequently a significant loss of atoms due to three-body recombination losses.

In Figure~\ref{6}, we examine the impact of initial momentum kicks on the FWHM, peak densities and atom loss. Each momentum kick requires a particular optimal power, since higher momentum kicks require larger powers to bring atoms to the same spot. For example, for a kick towards the focal plane with a magnitude of $2\hbar k$, the corresponding optimal power of $P_{2\hbar k}=4.188$ mW is required to focus at $z=z_f=0$. For larger momentum kicks, the powers needed for a focal plane of $z_f=0$ are $(P_{4\hbar k}, P_{6\hbar k}, P_{8\hbar k}, P_{10\hbar k}, P_{12\hbar k}, P_{14\hbar k}, P_{16\hbar k})=(4.372, 4.678, 5.106, 5.656, 6.329, 7.124, 8.041)$ mW. As indicated in Figure~\ref{6}(a), more powerful kicks reduce the widths of the deposited atoms along the $x$ direction. Increasing the momentum kick leads to slightly higher peak densities for repulsive BECs in regimes where the three-body recombination effects are insignificant [see Figure~\ref{6}(c)]. However, larger kicks give smaller peak densities for a small negative $a_s$, as seen in Figure~\ref{6}(b) for $a_s=-1a_0$.

Finally, in addition to the study of the accumulative atomic flux on a substrate during the focus process, it is also worth considering the resultant profile right at the moment of optimal focus at $z=z_f=0$, known as an instantaneous profile. This has been used in previous theoretical work to examine the structures predicted by the particle optics or classical trajectories approach \cite{20}. Figures~\ref{7}(a-c) illustrate the related outcomes of resolution, peak density and atom loss for a propagating condensate immediately at the time its center of mass reaches the center of the focusing plane, $z=0$. As above, it is assumed that the condensate has started its free propagation at $z_0=500~\mu$m and is exposed to a focusing potential comprising the parameters of $k = 2.01384 \times 10^4$ m$^{-1}$, $\sigma_z=100~\mu$m and $\Delta=200$ GHz. There appears a more rapid trend in FWHM over different ranges of scattering length [see Figure~\ref{7}(a)] when a momentum kick of $p=32\hbar k$ is applied compared to that of $p=64\hbar k$ and $p=128\hbar k$. The impact of increasing the scattering length on the structure resolution is more significant for lower kicks, especially in the regime of relatively low $a_s$ where the three-body losses are non-negligible. This is also the case for the scattering length dependence of the peak density [see Figure~\ref{7}(b)]. Although the trend remains steady with a gradual slope for $p=128\hbar k$, it undergoes a fluctuation around $a_s\sim1a_0$ for lower kicks (i.e. $p=64\hbar k$), which becomes steeper with a decrease in kick values as seen for $p=32\hbar k$. Turning to the atom number loss shown in Figure~\ref{7}(c), the amount of loss in high density regimes for lower kicks (e.g. $p=32\hbar k$) is considerably greater than higher kicks. In fact, the BEC with a slower longitudinal velocity is exposed to the field for a relatively longer time resulting in a larger atom loss whereas for higher velocities (i.e. $p=128\hbar k$), the BEC has less chance to interact with the field. Of the parameters we considered, $a_s=-1a_0$ and $p=128\hbar k$ gave the instantaneous profile with the best resolution (FWHM$_x$$\simeq$ 9 nm) and highest peak density ($2\times 10^6$ atoms/$\mu \text{m}^2$).

\section{\label{sec:level6}Numerical Simulations (GPE)}

In order to investigate the accuracy of our variational solutions, we compare with GPE numerical simulations. We numerically solved Eq.(\ref{e1}) with an embedded Runge-Kutta (ERK) scheme along with adaptive Fourier split-step size \cite{43}. We used the third and fourth orders [ERK4(3)] to deliver an estimation of the local error for adaptive step-size control purposes in the interaction picture. The initial condition for each simulation was the GPE groundstate, numerically determined via imaginary time evolution. Our simulations indicate that the variational method predicts the evolution of the BEC widths and density functions well. Figure~\ref{8} represents the top view of the deposited profile on the surface $z=0$ extracted by both the numerical GPE and variational approach when $a_s=-1a_0$ and with zero momentum kick. The GPE results for a variety of \textit{s}-wave scattering lengths, when the potential power is set to $P_{0\hbar k}$, are shown in Figure~\ref{9} as well as the corresponding variational calculations. There is a good agreement between the two approaches, especially in the interval of $a_s\leqslant 10a_0$. 

It is worth noting that for a BEC moving relatively slowly through the focusing potential, the dimensionality disruptions observed in the GPE simulations, which result from excitations, are negligible. Hence, in the slow regime, one can expect reasonable agreement between the focused profile distributions of the GPE and variational approaches. However, as higher momentum kicks are applied to the BEC, excitations during the focusing process become more significant. Figure~\ref{10} illustrates this by comparing the BEC density profiles along the $x$ axis as calculated from the variational approach and GPE simulations.
For $a_s=100a_0$ and $p=32\hbar k$ the GPE simulations show that significant fringes emerge, which are not captured in the variational methodology. The estimated resolutions are still in a close agreement, $(\Delta x)^{32\hbar k}_\text{var}=47.5$ nm and $(\Delta x)^{32\hbar k}_\text{GPE}=46.4$ nm. In the case of no momentum kick, both the GPE and variational distributions tend to a Gaussian profile (although the GPE curve has a longer tail than the variational one, see the solid green and dashed red curves). The tail in the GPE simulations reduces by increasing the momentum kick, which necessitates an increase in the potential power and intensity accordingly for the same focus spot. Again the estimated resolutions are in close agreement with $(\Delta x)^{0\hbar k}_\text{var}=130.1$ nm and  $(\Delta x)^{0\hbar k}_\text{GPE}=133.9$ nm.

\section{\label{sec:level7}Conclusion}

We have investigated the focusing dynamics of a $^{85}$Rb BEC with two-body interactions and three-body recombination losses due to a harmonic-shaped focusing optical potential. Using a variational technique, we derived the dynamical behavior of the peak densities, FWHMs, and atom number loss in the focusing regimes when tuning the scattering length over a large range. We showed that the inter-particle interaction can play an essential role in the deposited profile structures, which become more sensitive to smaller $a_s$ as well as its negative values. We conclude that high peak densities and small resolutions can be achieved in relatively attractive BECs. However, the three-body losses limit the maximum achievable density and, when sufficient, can disrupt the deposition quality. Since highly repulsive BECs are exposed to negligible amounts of atom loss in the focusing regime, applying a momentum kick to the BEC can always be beneficial to both the peak density and resolution of focused structures. Nevertheless, the scenario is completely different for the attractive BECs. In this case, although higher momentum kicks and powers may provide one with more resolution transparency, they can still destroy created peak densities. Last but not least, we inferred that to improve the profile resolution at the same focal plane, one needs to boost the momentum kick in accordance with the potential power, whereas for the same momentum kick, increasing the power leads to a focal plane displacement; this is optimally shifted along the longitudinal axis to an area above the center of the focusing potential.

Overall, we conclude that reaching nano-meter scale structures using a BEC source is achievable, which provides much higher profile resolution and peak density than those created by thermal atomic sources. Finally, we have demonstrated the power of using a variational methodology to examine a broad range of parameter space which would not be easily achievable using 3D GPE simulations. 

\section*{ACKNOWLEDGMENTS}

The authors would like to thank Nicholas P. Robins and Hans A. Bachor for useful discussions and feedback. Grateful acknowledgement is also extended to Timothy Senden for the financial support of the project. SSS was supported by an Australian Research Council Discovery Early Career Researcher Award (DECRA), project No. DE200100445.

\section*{Appendix}

Here we provide a detailed description of our analytical methods including a derivation of the Lagrangian Eq.(\ref{e12}) and the variational equations Eqs.(\ref{e18})-(\ref{e21}).

\subsection*{1. Derivation of Lagrangian function Eq.(\ref{e12})}
\label{subsec:a1}
Substituting Eq.(\ref{e9}) into Eq.(\ref{e8}) and simplifying give the Lagrangian density
\begin{equation}\tag{A1}
\begin{split}
\mathcal{L}(\mathbf{r}, t)=-\sum\limits_{r=x, y, z}|A|^2\exp{(-r^2/\mathcal{W}^2_r)}~~~~~~~~~~~~~~~~~~~~~~~\\
\times\Bigg\{\hbar\Bigg[\dot{\beta}_r r^2+\dot{\alpha_r}r+\dot{\phi}_r\Bigg]
+\frac{\hbar^2}{2m}\Bigg[\frac{r^2}{\mathcal{W}^4_r}+4\beta_r^2 r^2~~~~~\\
+4\beta_r\alpha_rr+\alpha_r^2\Bigg]-\frac{1}{2}|A|^2 u\Bigg\}~~~~~~~~~~~~~~~~~~~~~~~~~~~\\
-\frac{1}{2}m|A|^2\omega_{x}^2(t)x^2\exp{(-x^2/\mathcal{W}^2_x)}.~~~~~~~~~~~~~~~~~
\label{A1}
\end{split}
\end{equation}
Note that this excludes the three-body recombination loss term. Inserting Eq.(\ref{A1}) into $L(t)=\int d^3\mathbf{r}\ \mathcal{L}(\mathbf{r}, t)$ gives the Lagrangian \\
\begin{equation}\tag{A2}
\begin{split}
L(t)=-\sum\limits_{r=x, y, z}\frac{|A|^2\mathcal{W}_r\sqrt{\pi}}{2}\Bigg\{\hbar\dot{\beta}_r\mathcal{W}_r^2+2\hbar\dot{\phi}_r+\frac{\hbar^2}{2m}\Big(\frac{1}{\mathcal{W}_r^2}\Big)\\+\frac{\hbar^2}{2m}\big(4\beta_r^2\mathcal{W}_r^2\big)+\frac{\hbar^2}{2m}\big(2\alpha_r^2\big)+\frac{1}{2}m\omega_x^2(t)\mathcal{W}_x^2+\frac{g|A|^2}{\sqrt{2}}\Bigg\},
\label{A2}
\end{split}
\end{equation}
where the following three well-known Gaussian integrals have been used:
\begin{equation}\tag{A3}
\begin{split}
\int^{\infty}_{-\infty}\exp(-ar^2) \ dr=\sqrt{\frac{\pi}{a}},\\
\int^{\infty}_{-\infty}r\exp(-ar^2) \ dr=0,\\
\int^{\infty}_{-\infty}r^2\exp(-ar^2) \ dr=\frac{\sqrt{\pi}}{2a^{3/2}},
\label{A3}
\end{split}
\end{equation}
where $a$ is a constant.

\subsection*{2. Derivation of BEC Variational Equations}

Substituting Eq.(\ref{e12}) into Eq.(\ref{e17}) and solving for $q_i=\mathcal{A}, \mathcal{W}_r, \alpha_r, \beta_r, \phi_r$ gives the following respective equations:
\begin{equation}\tag{A4}
\begin{split}
\sum\limits_{r=x, y, z}\Bigg\{\frac{\hbar\dot{\beta}_r\mathcal{W}_r^2}{2}+\hbar\dot\phi_r+\frac{\hbar^2}{m}\Big(\frac{1}{4\mathcal{W}_r^2}\Big)+\frac{\hbar^2}{m}(\beta_r^2\mathcal{W}_r^2)+\\
+\frac{\hbar^2}{2m}\alpha_r^2+\frac{g|A|^2}{\sqrt{2}}\Bigg\}+\frac{1}{4}m\omega_x^2(t)\mathcal{W}_x^2=0,\quad\quad
\label{A4}
\end{split}
\end{equation}
\begin{equation}\tag{A5}
\begin{split}
\sum\limits_{r=x, y, z}\Bigg\{\frac{3}{2}\hbar\dot{\beta}_r\mathcal{W}_r^2+\hbar\dot\phi_r-\frac{\hbar^2}{m}\Big(\frac{1}{4\mathcal{W}_r^2}\Big)+\frac{\hbar^2}{m}(3\beta_r^2\mathcal{W}_r^2)\\
+\frac{\hbar^2}{2m}\alpha_r^2+\frac{g|A|^2}{2\sqrt{2}}\Bigg\}+\frac{3}{4}m\omega_x^2(t)\mathcal{W}_x^2=0,\quad\quad
\label{A5}
\end{split}
\end{equation}
\begin{equation}\tag{A6}
\sum\limits_{r=x, y, z}\Bigg\{\frac{\hbar^2}{m}|A|^2\mathcal{W}_r\alpha_r\Bigg\}=0,
\label{A6}
\end{equation}
\begin{equation}\tag{A7}
\begin{split}
\sum\limits_{r=x, y, z}\Bigg\{\hbar\frac{d}{dt}\big(|A|^2\mathcal{W}_r^3\big)-\frac{\hbar^2}{m}(4|A|^2\beta_r\mathcal{W}_r^3)+\frac{K|A|^6\mathcal{W}_r^3}{9\sqrt{3}}\Bigg\}=0,
\label{A7}
\end{split}
\end{equation}
\begin{equation}\tag{A8}
\begin{split}
\sum\limits_{r=x, y, z}\Bigg\{\hbar\frac{d}{dt}\big(|A|^2\mathcal{W}_r\big)+\frac{K|A|^6\mathcal{W}_r}{3\sqrt{3}}\Bigg\}=0.
\label{A8}
\end{split}
\end{equation}

According to the normalization condition $N(t)=\int_{-\infty}^{+\infty} \psi\psi^*\ dr$, and Eq.(\ref{e9}), we can write
\begin{equation}\tag{A9}
\frac{dN(t)}{dt}=\sqrt{\pi}\frac{d}{dt}\big(|A|^2\mathcal{W}_r\big).
\label{A9}
\end{equation}
Combining Eq.(\ref{A7}), Eq.(\ref{A8}), and using Eq.(\ref{e10}) gives the associated differential equation for the change of atom number in the condensate
\begin{equation}\tag{A10}
\frac{dN(t)}{dt}=-\frac{KN^3}{3\sqrt{3}\ \hbar\pi\mathcal{W}_r^2}.
\label{A10}
\end{equation}
We then merge Eq.(\ref{A4}) and Eq.(\ref{A5}) to acquire
\begin{equation}\tag{A11}
\begin{split}
\sum\limits_{r=x, y, z}\Bigg\{\hbar\dot{\beta}_r\mathcal{W}_r^2+\frac{\hbar^2}{m}\Big(2\beta_r^2\mathcal{W}_r^2-\frac{1}{2\mathcal{W}_r^2}\Big)-\frac{g|A|^2}{2\sqrt{2}}\Bigg\}\\
+\frac{1}{2}m\omega_x^2(t)\mathcal{W}_x^2=0.~~~~~~~~~~~~~~~~~~~~~~~~~~~~~~~~~
\label{A11}
\end{split}
\end{equation}
Considering the amplitude $|A|$ via Eq.(\ref{e10}), this implies that
\begin{equation}\tag{A12}
\begin{split}
\dot{\beta}_x=\Big(\frac{\hbar}{2m}\Big)\frac{1}{\mathcal{W}_x^4}-\Big(\frac{2\hbar}{m}\Big)\beta_x^2-\Big(\frac{m}{2\hbar}\Big)\omega_x^2(t)\\
+\frac{gN}{2(2\pi)^{3/2}\hbar\mathcal{W}_x^3\mathcal{W}_y\mathcal{W}_z},~~~~~~~~~~~~~~~~~~~
\label{A12}
\end{split}
\end{equation}
\begin{equation}\tag{A13}
	\begin{split}
		\dot{\beta}_y=\Big(\frac{\hbar}{2m}\Big)\frac{1}{\mathcal{W}_y^4}-\Big(\frac{2\hbar}{m}\Big)\beta_y^2+\frac{gN}{2(2\pi)^{3/2}\hbar\mathcal{W}_x\mathcal{W}_y^3\mathcal{W}_z},
		\label{A13}
	\end{split}
\end{equation}
\begin{equation}\tag{A14}
	\begin{split}
		\dot{\beta}_z=\Big(\frac{\hbar}{2m}\Big)\frac{1}{\mathcal{W}_z^4}-\Big(\frac{2\hbar}{m}\Big)\beta_z^2+\frac{gN}{2(2\pi)^{3/2}\hbar\mathcal{W}_x\mathcal{W}_y\mathcal{W}_z^3}.
		\label{A14}
	\end{split}
\end{equation}
Similarly mixing Eq.(\ref{A7}) and Eq.(\ref{A8}), gives
\begin{equation}\tag{A15}
\sum\limits_{r=x, y, z}\Bigg\{\hbar\dot{\mathcal{W}}_r-\frac{2\hbar^2}{m}(\beta_r\mathcal{W}_r)-\frac{K|A|^4\mathcal{W}_r}{9\sqrt{3}}\Bigg\}=0.
\label{A15}
\end{equation}
Again using Eq.(\ref{e10}), this gives the evolution of the $x$ component width

\begin{equation}\tag{A16}
\dot{\mathcal{W}}_x=\frac{2\hbar}{m}(\beta_x\mathcal{W}_x)+\frac{KN^2}{\sqrt{3}(3\pi)^3\hbar\mathcal{W}_x\mathcal{W}_y^2\mathcal{W}_z^2}.
\label{A16}
\end{equation}
Taking the second derivative of (\ref{A16}) with respect to time for $x$-dependent variables gives
\begin{equation}\tag{A17}
\begin{split}
\ddot{\mathcal{W}}_x=\frac{2\hbar}{m}(\dot{\beta_x}\mathcal{W}_x+\beta_x\dot{\mathcal{W}_x})+\frac{2\dot{N}NK}{\sqrt{3}(3\pi)^3\hbar\mathcal{W}_x\mathcal{W}_y^2\mathcal{W}_z^2}\\
-\frac{KN^2\dot{\mathcal{W}_x}}{\sqrt{3}(3\pi)^3\hbar\mathcal{W}_x^2\mathcal{W}_y^2\mathcal{W}_z^2}.~~~~~~~~~~~~~~~~~~~~~~~~~~~~
\label{A17}
\end{split}
\end{equation}
Using the associated equation for atom number variation in terms of all $x$, $y$, and $z$ components derived from Eq.(\ref{A9}),
\begin{equation}\tag{A18}
\frac{dN(t)}{dt}=-\frac{KN^3}{9\sqrt{3}\pi^3\hbar\mathcal{W}_x^2\mathcal{W}_y^2\mathcal{W}_z^2},
\label{A18}
\end{equation}
as well as Eq.(\ref{A12}) and Eq.(\ref{A16}), the variational equation for the condensate width along the $x$ axis is obtained Eq.(\ref{e18}). Repeating the same procedure for the $y$ and $z$ components gives the variational Eqs.(\ref{e19}) and (\ref{e20}).


\bibliography{Ref_1,Ref_2,Ref_3,Ref_4,Ref_5,Ref_6,Ref_7,Ref_8,Ref_9,Ref_10,Ref_11,Ref_12,Ref_13,Ref_14,Ref_15,Ref_16,Ref_17,Ref_18,Ref_19,Ref_20,Ref_21,Ref_22,Ref_23,Ref_24,Ref_25,Ref_26,Ref_27,Ref_28,Ref_29,Ref_30,Ref_31,Ref_32,Ref_33,Ref_34,Ref_35,Ref_36,Ref_37,Ref_38,Ref_39,Ref_40,Ref_41,Ref_42,Ref_43,Ref_44,Ref_45,Ref_46,Ref_47,Ref_48,Ref_49,Ref_50,Ref_51,Ref_52,Ref_53,Ref_54,Ref_55,Ref_56,Ref_57,Ref_58,Ref_59,Ref_60,Ref_61,Ref_62,Ref_63,Ref_64,Ref_65,Ref_66,Ref_67,Ref_68,Ref_69,Ref_70,Ref_71,Ref_72,Ref_73,Ref_74,Ref_75,Ref_76,Ref_77,Ref_78,Ref_79,Ref_80,Ref_81,Ref_82,Ref_83,Ref_84,Ref_85,Ref_86,Ref_87,Ref_88,Ref_89,Ref_90,Ref_91,Ref_92,Ref_93,Ref_94,Ref_95,Ref_96,Ref_97,Ref_98,Ref_99,Ref_100,Ref_101,Ref_102,Ref_103,Ref_104,Ref_105,Ref_106,Ref_107,Ref_108,Ref_109}
\end{document}